\documentclass[floats,aps,amsmath,amssymb,twocolumn,longbibliography]{revtex4-2} 
\usepackage[utf8]{inputenc}
\usepackage{amsmath}
\usepackage{array}
\usepackage{graphics}
\usepackage{multirow}
\usepackage{cleveref}

\usepackage{natbib}
\usepackage{graphicx}
\usepackage{xcolor}

\begin{document}

\title{Improved accuracy for decoding surface codes with matching synthesis}
\author{Cody Jones}
\email{ncodyjones@google.com}
\affiliation{Google Quantum AI, Santa Barbara, CA 93117, USA}

\begin{abstract}
    We present a method, called matching synthesis, for decoding quantum codes that produces an enhanced assignment of errors from an ensemble of decoders.  We apply matching synthesis to develop a decoder named Libra, and show in simulations that Libra increases the error-suppression ratio $\Lambda$ by about $10\%$.  Matching synthesis takes the solutions of an ensemble of approximate solvers for the minimum-weight hypergraph matching problem, and produces a new solution that combines the best local solutions, where locality depends on the hypergraph.  We apply matching synthesis to an example problem of decoding surface codes with error correlations in the conventional circuit model, which induces a hypergraph with hyperedges that are local in space and time.  We call the matching-synthesis decoder Libra, and in this example the ensemble consists of correlated minimum-weight matching using a different hypergraph with randomly perturbed error probabilities for each ensemble member.  Furthermore, we extend matching synthesis to perform summation of probability for multiple low-weight solutions and at small computational overhead, approximating the probability of an equivalence class; in our surface code problem, this shows a modest additional benefit.  We show that matching synthesis has favorable scaling properties where accuracy begins to saturate with an ensemble size of 60, and we remark on pathways to real-time decoding at near-optimal decoding accuracy if one has an accurate model for the distribution of errors.
\end{abstract}

\maketitle

\section{Introduction}
A fault-tolerant quantum computer requires real-time decoding of the parity checks that detect errors during the computation~\cite{Terhal2015}.  Accuracy of decoding is important as well, since improvements in accuracy can reduce the resource overhead for error correction~\cite{steane2003,aliferis2006,fowler2012largescale,jones2012arch}.  Hence decoding algorithms that achieve higher accuracy while being efficient to run in real time are desirable for making useful quantum computers.

Many works have studied accurate decoding for surface codes.  If accuracy improvements are relative, then the historical baseline~
\cite{dennis2002topological} is Edmonds' Blossom algorithm~\cite{edmonds1965} for minimum-weight perfect matching (where often the decoder itself is known as MWPM).  The observation that correlations between the $X$ and $Z$ lattices (or primal and dual in the original work~\cite{kitaev1998surfacecode}) could be exploited by edge reweighting~\cite{hutter2014} led to improved accuracy with correlated MWPM~\cite{fowler2013optimal,Paler2023pipelinedcorrelated,higgott2023improved,shutty2024harmony}.
More recently, a mapping between surface and color codes~\cite{benhemou2023colorcode} converts the problem into finding a good decoder for color codes.  Accuracy in matching can also be improved with belief propagation~\cite{higgott2023bp} or accounting for path degeneracy with multi-path summation~\cite{criger2018multipath}.

Whereas correlated matching is heuristic with error correlations, there are other decoding algorithms that attempt to compute optimal solutions.  Belief propagation~\cite{old2023,chen2024bp} can achieve high accuracy but sometimes fails to converge (Ref.~\cite{higgott2023bp} falls back to matching).  Tensor-network decoders appear to approach optimal accuracy for 2D networks (noiseless syndrome measurement)~\cite{bravyi2014efficient,chubb2021statistical,chubb2021general}, but contracting a 3D tensor network, required for fault tolerance in surface codes, is believed to be intractable beyond small code distances~\cite{piveteau2023tensor,google2023suppressing, bohdanowicz2022quantum}.  Likewise, integer-program decoders have exponential runtime in the worst case~\cite{feldman2005using,landahl2011fault}, though the average runtime for below-threshold error rates could be more favorable; linear-program decoders~\cite{livontobel,fawzihypergraph} may also enable circumventing intractable runtimes.

This work introduces a decoder called Libra that utilizes ensembling, where a collection of decoders produce differing decoding solutions, which are analyzed collectively.  Ensembling for decoding has appeared in machine-learning decoders~\cite{sheth2020neural,bausch2023learning} and matching with the Harmony decoder~\cite{shutty2024harmony}.  The Harmony decoder, developed by colleagues at the same institution, was an inspiration for how to identify improving cycles in problem-independent way, as opposed to local search that would introduce complexity that depends on problem structure.  Both Harmony and Libra access an ensemble of decoders; Harmony makes fewer demands on the ensemble output (majority voting or selecting the global-minimum weight), whereas Libra can synthesize the best local solutions across the ensemble, if the ensemble can produce an assignment of errors.

We remark that the matching synthesis method appears new to the author, but the procedure seems to apply to other problems in optimization beyond decoding quantum codes.  Hence this method or something similar may already appear in a different field, under a different name.  It was not found in a literature search, though this would be limited by the author's knowledge of terminology in other fields.  The use of a random ensemble and combining pieces of solutions resembles local search in simulated annealing~\cite{lin1965,lin1973,martin1991} or ``crossover'' in genetic algorithms~\cite{mahfoud1995,muhlenbein1988,manzoni2020}, but the similarity ends there; our method is not local search or a genetic algorithm.  If a closer match to this method is found in prior work, the manuscript will be updated accordingly.

\section{Decoding from Minimum-Weight Hypergraph Perfect Matching}
\label{sec:MWHPM_def}
The minimum-weight hypergraph perfect matching (MWHPM) problem is a generalization of the more commonly encountered minimum-weight perfect matching problem~\cite{edmonds1965}.  However, whereas the latter can be solved optimally in polynomial time, MWHPM is an NP-hard problem~\cite{hastad1999}.  Hence the best we can hope for in a new approach to this problem is to improve the trade-off between computational time and approximation to optimal accuracy.  Let us first define the MWHPM problem and describe how this problem applies to decoding quantum codes.

Define a weighted hypergraph as $G = (V, E)$ where $V$ is a set of vertices and $E$ is a set of hyperedges.  Each hyperedge $e \in E$ has some list of vertices $v(e)$, where the list size is any positive integer (contrast this with a graph, where the size must be two for every edge) and every vertex in the list is in $V$.  Furthermore, each hyperedge $e$ has weight $w(e)$ that in general can be any real number, though we restrict our attention to $w(e) \ge 0$ in this work.  Let us furthermore represent the sum of edge weights for a list of hyperedges $\mathbf{e} = \{e_i\}$ as
\begin{equation}
    w(\mathbf{e}) = \sum_{e_i \in \mathbf{e}} w(e_i).
\end{equation}

In mapping to decoding stabilizer quantum codes~\cite{gottesman1998heisenberg}, each vertex is a parity check, and each hyperedge corresponds to an error channel.  The vertices $v(e)$ are the parity checks that flip when the error $e$ occurs.  There are different approaches to selecting the hyperedge weights, but we will motivate a typical one.  If errors are independent binary channels with probability $p(e)$ (i.e. weighted coin flip for each error occurring or not), then setting $w(e) = \log [(1 - p(e))/p(e)]$ creates an equivalence between minimizing a sum of hyperedge weights and maximizing posterior probability over error events consistent with the syndrome~\cite{hutter2014, bravyi2014efficient,chubb2021statistical, shutty2024harmony}, as described below.  Because each hyperedge $e$ corresponds to a specific error occurring, we will interchangeably refer to $e$ as an error as well.  Likewise, we will say $\mathbf{e}$, a list of hyperedges, is also a configuration of errors.

The MWHPM problem, generalized for our purposes, is defined as follows.  For a weighted hypergraph, one is given a ``syndrome'' $S$ that is a list of vertices.  In decoding quantum codes, the syndrome will be parity checks that have flipped due to errors.  Furthermore, let us define a syndrome function $s(e) = v(e)$ for a single hyperedge, and for a list of hyperedges $\mathbf{e} = \{e_i\}$ 
\begin{equation}
    s(\mathbf{e}) = \bigoplus_{e_i \in \mathbf{e}} s(e_i),
    \label{eqn:syndrome}
\end{equation}
where $s(e_1) \oplus s(e_2)$ means symmetric difference of the lists $s(e_1)$ and $s(e_2)$, or equivalently mod-2 parity.  Hence, the MWHPM optimization problem is finding $\mathbf{e}$ that solves
\begin{equation}
    \min w(\mathbf{e}) \; \mathrm{s.t.} \; s(\mathbf{e}) = S.
    \label{eqn:weight_optimization}
\end{equation}
Said in words, all configurations of hyperedges that satisfy $s(\mathbf{e}) = S$ correspond to possible Pauli-error patterns consistent with the observed parity-check violations, and we seek the error pattern that minimizes the weight sum $w(\mathbf{e})$ because this event will be the most probable.  For simplicity, we call any solution $\mathbf{e}$ that satisfies $s(\mathbf{e}) = S$ a ``matching for $S$''.

The generalization to MWHPM used here is in the meaning of ``perfect matching'', implied by the $\oplus$ in Eqn.~(\ref{eqn:syndrome}).  Here we take perfect matching to mean that the parity of hyperedges incident to a given vertex $u$ must be odd for $u \in S$ and must be even for $u \notin S$.  In the traditional formulation of perfect matching for graphs~\cite{edmonds1965}, there is no syndrome (or one could say the syndrome is implicitly all vertices), and the number of incident edges to every vertex must be one, instead of any odd number.

To complete the connection between MWHPM and decoding for quantum codes, we additionally associate an ``observable list'' $o(e)$ with every hyperedge $e$.  This does not appear in the optimization problem, but it does determine the result of decoding once a solution to the optimization problem has been chosen.  For the quantum code, make some arbitrary choice of logical operators $\{O_i\}$ (which are Pauli strings in stabilizer codes) that propagate through the error correction circuit.  For each error event, $o(e)$ is the list of operators in $\{O_i\}$ that anticommutes with $e$ (i.e. error $e$ occurring would flip the observed value of $O_i$).  For a set of hyperedges we can say that
\begin{equation}
    o(\mathbf{e}) = \bigoplus_{e_i \in \mathbf{e}} o(e_i),
\end{equation}
where again $\oplus$ means symmetric difference.  Equivalently, if we track observable lists as bitstrings with a 1 in position $i$ for each $O_i$, then the operation $\oplus$ means bitwise XOR.  

These logical observables are important for two reasons.  First, all error patterns with the same observable list define an equivalence class~\cite{hutter2014,bravyi2014efficient}.  Second, when decoding is employed in a fault-tolerant computation, the observable list for the chosen error configuration determines the outcome of logical-qubit measurements.

\section{Matching Synthesis}
\label{sec:matching_synthesis}
Matching synthesis is a procedure for taking two or more solutions to the MWHPM problem, described in the previous section, and producing a new solution that is in general better than any of the inputs (and at least as good as the best input).  For two distinct MWHPM solutions, the procedure is:
\begin{enumerate}
    \item compute the symmetric difference (SD) of the assigned errors,
    \item separate the SD into pieces with null syndrome (``cycles''),
    \item determine the signed weight of each cycle, and
    \item form a new ``synthetic matching'' that incorporates only the cycles that lower weight.
\end{enumerate}
The rest of this section elaborates on the synthesis process, and later sections describe how this can be used in decoding quantum codes.

Following the nomenclature of Section~\ref{sec:MWHPM_def}, suppose that for decoding a syndrome $S$, there are candidate matchings $\mathbf{e}$ and $\mathbf{f}$: $s(\mathbf{e}) = s(\mathbf{f}) = S$.  Let $\mathbf{d} = \mathbf{e} \oplus \mathbf{f}$ be the symmetric difference of these two solutions; namely, cancel out the errors assigned in both configurations and keep the errors that appear in only one configuration.  By linearity, $s(\mathbf{d}) = s(\mathbf{e}) \oplus s(\mathbf{f}) = \emptyset$, the null syndrome.

Any set of hyperedges $\mathbf{c}$ with null syndrome has a property that is key to this work.  Let us label any such $\mathbf{c}$ a ``cycle'', where we will refine the definition below after motivating its use.  For any error configuration $\mathbf{e}$ and any cycle $\mathbf{c}$, then $s(\mathbf{e} \oplus \mathbf{c}) = s(\mathbf{e})$.  Hence, given one decoding solution $s(\mathbf{e}) = S$, we can use cycles to generate other solutions.  If we are clever about which cycles to incorporate, then we can produce new solutions with lower weight, $w(\mathbf{e} \oplus \mathbf{c}) < w(\mathbf{e})$ and improve our approximate solution to MWHPM.

\begin{figure}
  \centering
      \includegraphics[width=0.9\columnwidth]{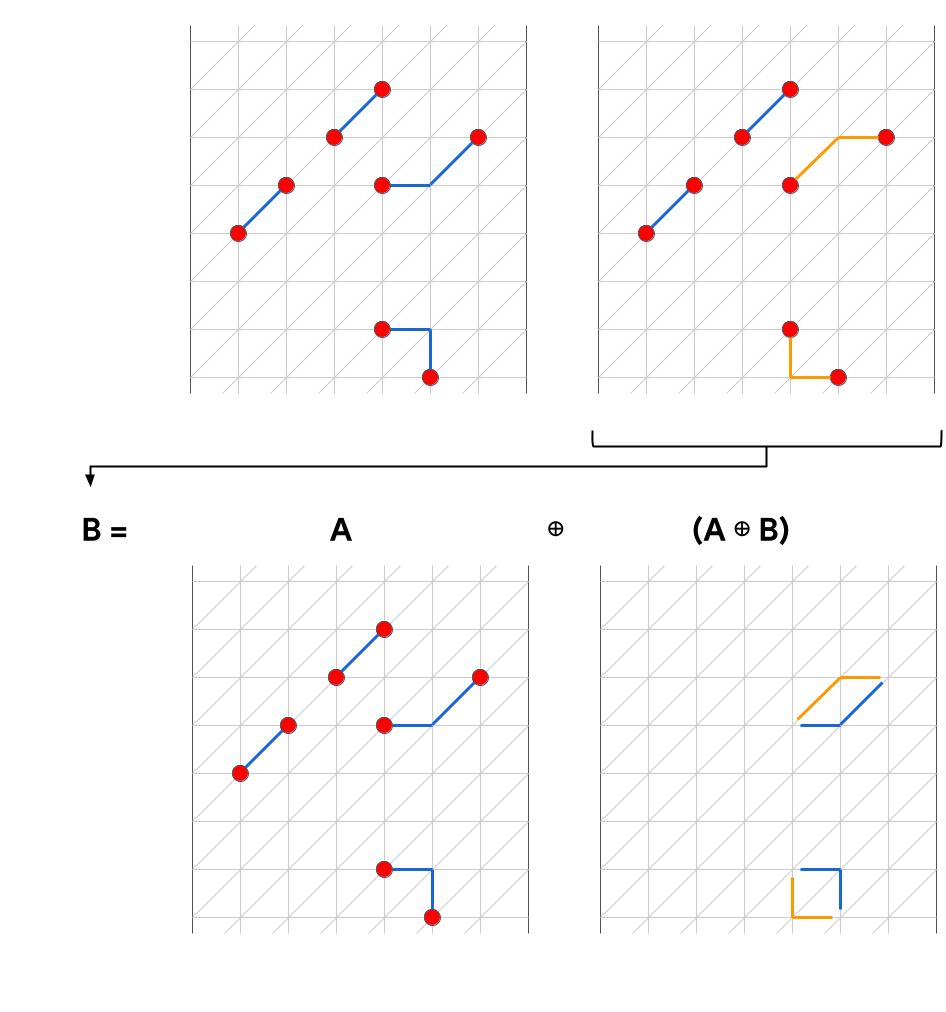}
  \caption{Example showing that two matchings ($A$ and $B$, top) for the same syndrome can be related by flipping two cycles (bottom), where the cycles are collectively the symmetric difference $\mathbf{A} \oplus \mathbf{B}$.}  
  \label{fig:symmetric_diff}
\end{figure}

Let us be explicit about what we mean by ``cycle'' in our extended version of a hypergraph.  For our purposes, a cycle is any hyperedge list $\mathbf{c}$ such that $s(\mathbf{c}) = \emptyset$ and $o(\mathbf{c}) = \emptyset$.  Setting aside logical observables, this definition subsumes the traditional meaning of cycles on graph.  It also includes any configuration of hyperedges such every vertex has an even number of incident hyperedges.  The logical observables are furthermore necessary because they define equivalence classes, and we require that given a cycle $\mathbf{c}$, $o(\mathbf{e}) = o(\mathbf{e} + \mathbf{c})$ for all $\mathbf{e}$, such that incorporating a cycle does not change the equivalence class for a given solution.  An example using edges in a normal graph, which is easier to visualize than a hypergraph, is shown in Fig.~\ref{fig:symmetric_diff}.

We want to make the best (i.e. lowest weight) matching that we can for a given syndrome $S$, and our approach will be to improve an initial solution $\mathbf{e}$ by combining with one or more cycles.  To proceed, we define a ``relative weight'' function for cycles:
\begin{equation}
    w(\mathbf{c} | \mathbf{e}) = w(\mathbf{c} \oplus \mathbf{e}) - w(\mathbf{e}).
    \label{eqn:relative_weight_full}
\end{equation}
While it might seem like we are going in circles, this relative weight can be expressed as
\begin{equation}
    w(\mathbf{c} | \mathbf{e}) = w(\mathbf{c} \backslash \mathbf{e}) - w(\mathbf{c} \cap \mathbf{e}),
    \label{eqn:relative_weight}
\end{equation}
where $\mathbf{c} \backslash \mathbf{e}$ means ``hyperedges in $\mathbf{c}$ and not in $\mathbf{e}$'' and $\mathbf{c} \cap \mathbf{e}$ is the interection, ``hyperedges in both $\mathbf{c}$ and $\mathbf{e}$''.  The interpretation is straightforward: in going from $\mathbf{e} \rightarrow \left(\mathbf{e} \oplus \mathbf{c}\right)$, we remove the hyperedges in $\mathbf{c} \cap \mathbf{e}$ and subtract their summed weight, and we we add the hyperedges in $\mathbf{c} \backslash \mathbf{e}$ and add their weight.  The formulation in Eqn.~(\ref{eqn:relative_weight}) shows that we can compute the relative weight of the cycle with complexity that scales with the cycle size, not the size of the matching as one might think from Eqn.~(\ref{eqn:relative_weight_full}). The relative weight is useful because it tells us how much a matching solution improves by combining with a cycle.  

\begin{figure}
  \centering
      \includegraphics[width=0.9\columnwidth]{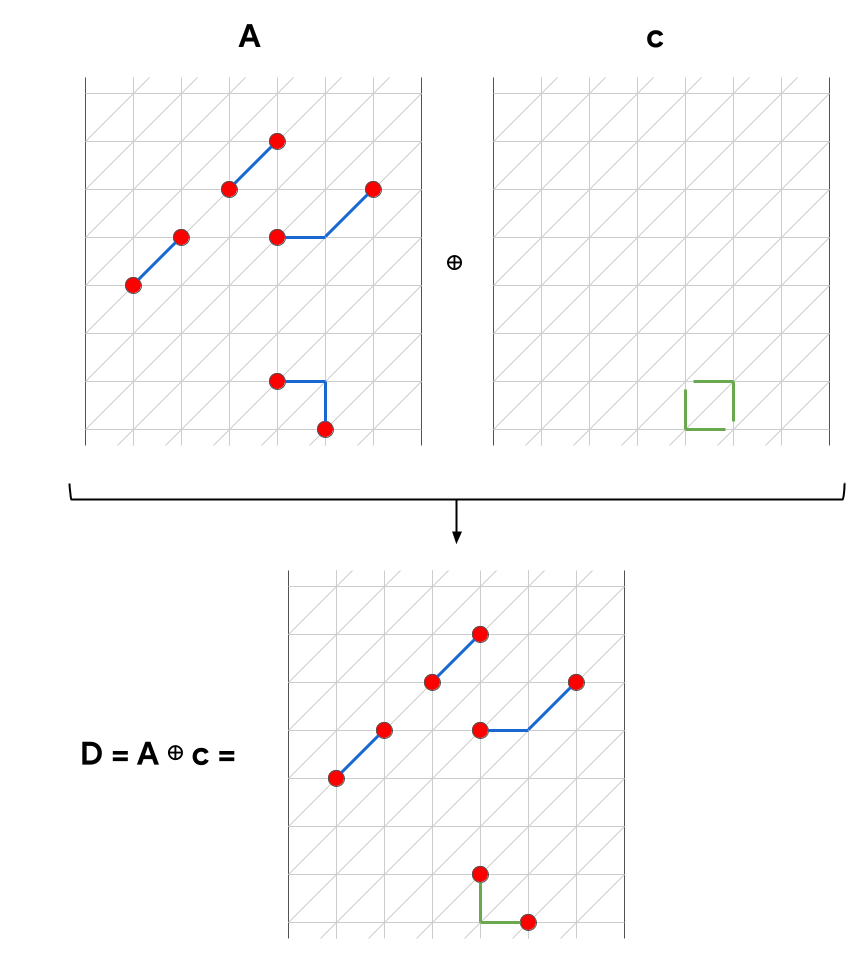}
  \caption{Continuing from the example in Fig.~\ref{fig:symmetric_diff}, suppose that just one of the cycles $\mathbf{c}$ from the lower-right of Fig.~\ref{fig:symmetric_diff} has negative relative weight.  We can produce a new ``synthetic'' matching $\mathbf{D} = \mathbf{A} \oplus \mathbf{c}$ that has lower weight than either of the source matchings $\mathbf{A}$ or $\mathbf{B}$.} 
  \label{fig:synthesis}
\end{figure}

So we want to find cycles with negative relative weight, $w(\mathbf{c} | \mathbf{e}) < 0$, because this would mean  $w(\mathbf{c} \oplus \mathbf{e}) < w(\mathbf{e})$.  If we take two matchings $\mathbf{e}$ and $\mathbf{f}$, compute the symmetric difference $\mathbf{d} = \mathbf{e} \oplus \mathbf{f}$, and compute $w(\mathbf{d} | \mathbf{e})$, we are left with the uninspiring result that we are just picking the lower weight matching of the two.  Now we reach the core result of matching synthesis.  Instead of computing the weight of the entire symmetric difference, we break $\mathbf{d}$ into smaller cycles where possible,
\begin{equation}
    \mathbf{d} = \bigoplus_i \mathbf{d}_i \;\; \mathrm{s.t.} \;\; s(\mathbf{d}) = s(\mathbf{d}_i) = \emptyset,
\label{eqn:diff_decomp}
\end{equation}
where each $\mathbf{d}_i$ is a subset of $\mathbf{d}$.  We then compute $w(\mathbf{d}_i | \mathbf{e})$ for each subset, and filter out the ones that have $o(\mathbf{d}_i) \ne \emptyset$ or $w(\mathbf{d}_i | \mathbf{e}) \ge 0$.  What remains are cycles $\{\mathbf{c}_i\}$ of negative relative weight.  We can produce a new matching
\begin{equation}
    \mathbf{g} = \mathbf{e} \oplus \left(\bigoplus_{\mathbf{c} \in \mathcal{N}} \mathbf{c}\right),
\end{equation}
where the set of negative-relative-weight cycles is $\mathcal{N} = \{\mathbf{c} : \mathbf{c} \in \mathbf{d}, w(\mathbf{c} | \mathbf{e}) < 0\}$.  Matching $\mathbf{g}$ is derived from $\mathbf{e}$ and $\mathbf{f}$, and it will equal $\mathbf{e}$ when $\mathcal{N} = \emptyset$ and equal $\mathbf{f}$ when $\mathcal{N} = \mathbf{d}$.  However, something interesting happens when $\mathcal{N}$ contains some of, but not all of, the cycles in $\mathbf{d}$: in this case, $\mathbf{g}$ is an entirely new ``synthetic'' matching, and it has lower weight than either $\mathbf{e}$ or $\mathbf{f}$.  This is depicted with a visual example in Fig.~\ref{fig:synthesis}.

Splitting a given $\mathbf{d}$ with $s(\mathbf{d}) = \emptyset$ into pieces $\{\mathbf{d_i}\}$ with $s(\mathbf{d_i}) = \emptyset$ can be done in more than one way.  The method we employ here is to simply to break $\mathbf{d}$ into connected components, where two hyperedges are connected when they touch a common vertex.  This can be computed efficiently using standard techniques in graph search.  It results in subsets $\mathbf{d_i}$ that satisfy Eqn.~(\ref{eqn:diff_decomp}) by being non-overlapping; note however than Eqn.~(\ref{eqn:diff_decomp}) does not require $\{\mathbf{d}_i\}$ to be non-overlapping.  Using connected components does not guarantee the pieces are as small as possible; for example, if there are two cycles that touch at one mutual vertex, they will form a single connected component.  We find that isolating cycles by connected components works well for decoding surface codes, as shown in Section~\ref{sec:simulations}, and we leave the more general problem of cycle decomposition, which can be done with cycle-finding algorithms, to future work.

There is one final step to close the loop on matching synthesis, and that is addressing the observable lists associated with hyperedges.  When we separate a symmetric difference $\mathbf{d}$ into pieces according to Eqn.~(\ref{eqn:diff_decomp}), such as by connected components, sometimes pieces in $\{\mathbf{d}_i\}$ will be logical operators.  Using our previous definitions, any $\mathbf{d}_i$ with $s(\mathbf{d}_i) = \emptyset$ and $s(\mathbf{d}_i) \ne \emptyset$ is a logical operator.  Although they are not cycles, logical operators will be useful in two ways for improving our decoding solutions.  First, using one solution $\mathbf{e}$, we can generate another solution in a \emph{different} equivalence class via $\mathbf{e} \oplus \mathbf{d_i}$, because $o(\mathbf{e} \oplus \mathbf{d_i}) = o(\mathbf{e}) \oplus o(\mathbf{d_i})$.  Second, for two pieces $\mathbf{d}_i$ and $\mathbf{d}_j$ with $s(\mathbf{d}_i) = s(\mathbf{d}_j) = \emptyset$ and $o(\mathbf{d}_i) = o(\mathbf{d}_j)$, we can make a cycle $\mathbf{d}_i \oplus \mathbf{d}_j$, because $o(\mathbf{d}_i \oplus \mathbf{d}_j) = o(\mathbf{d}_i) \oplus o(\mathbf{d}_j) = \emptyset$.  In some cases, adding cycles from a pair of equivalent logical operators, which may not be found by the connected-components method, can yield an improving cycle.

While matching synthesis improves a solution to the MWHPM problem by incorporating cycles of negative relative weight, what can we make of cycles with positive relative weight?  One application, which will we show can improve decoding for surface codes, is to record cycles with small positive (or zero) relative weight.  Each cycle can generate a new matching, and if the relative weight is small, this will be another approximate solution, with weight that is not the best known, but close to it.  Of course, if there are cycles with relative weight zero, they produce alternative solutions with smallest weight found.  If these small-weight cycles are non-overlapping, they generate many good solutions to the MWHPM problem: $n$ non-overlapping cycles encodes $2^n$ distinct solutions.  We show later that the Libra decoder can efficiently compute the sum of probability over all $2^n$ configurations in time linear in $n$, and also describe how it handles cases where these small-positive cycles overlap.  This is reminiscent of multi-path summation~\cite{criger2018multipath} or the way Markov-chain~\cite{hutter2014}, tensor-network~\cite{ bravyi2014efficient,chubb2021statistical, shutty2024harmony} or belief-propagation~\cite{old2023,higgott2023bp, chen2024bp} decoders account for multiple error configurations in an equivalence class.

\section{Libra decoder}
\label{sec:libra_decoder}
We apply matching synthesis to make a new decoder that we call Libra.  Libra works by using an ensemble of decoders to generate multiple distinct matchings, and synthesize from them better matchings.  Libra iteratively improves two or more matchings, one for each equivalence class represented.  For example, in the simulations of Section~\ref{sec:simulations}, there are two equivalence classes for a logical memory experiment (e.g. in an $X$-basis memory experiment, $X$ logical operators act trivially, so there are two equivalence classes and not four).

\begin{figure*}
  \centering
      \includegraphics[width=\textwidth]{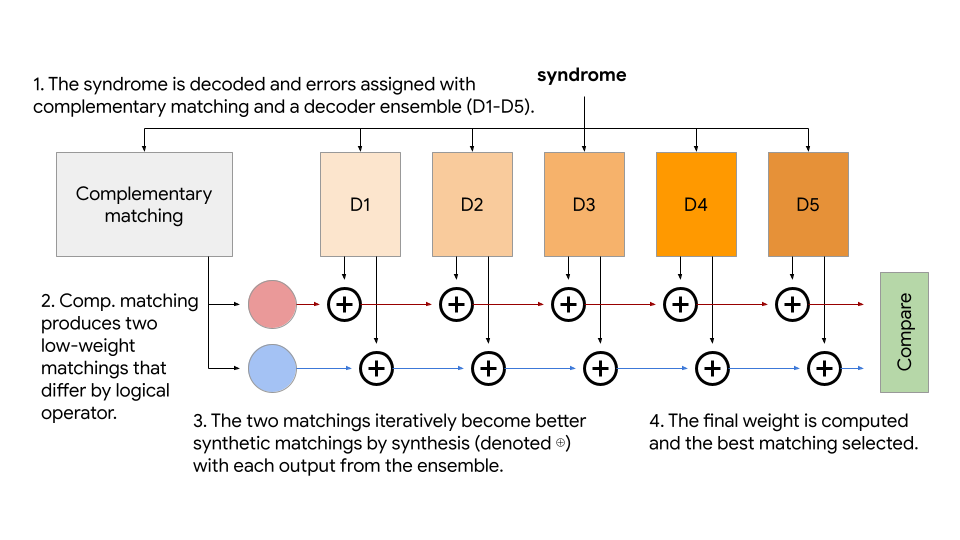}
  \caption{Architecture of the Libra decoder in this work.  The diagram depicts 5 decoders in $D1$--$D5$ in the ensemble, but the ensemble size can be arbitrary.  Each $\oplus$ operation in this diagram stands for matching synthesis (Sec.~\ref{sec:matching_synthesis}) between the two matchings indicated by arrows pointing into the $\oplus$. The arrow coming out of the $\oplus$ indicates the result of incorporating the cycles of negative relative weight (relative to the matching from the horizontal arrow, which originates with the complementary matching step).  As described in the text, when recording small-positive cycles, the sequence of synthesis stages in Step 3 is performed twice.  In the final step, a comparison is made between two matchings after incorporating improving cycles from the ensemble.}  
  \label{fig:libra_architecture}
\end{figure*}

The architecture of Libra as implemented in this work is shown in Fig.~\ref{fig:libra_architecture}.  To initialize a representative for each equivalence class, we first perform complementary matching~\cite{hutter2014,gidney2023yoked,bombin2024,meister2024,smith2024}, using correlated MWPM on the unperturbed error hypergraph.  Correlated MWPM returns two ``edge'' matchings for the $X$ and $Z$ graphs, which can be converted to the most-probable assignment of hyperedges within the original hypergraph that contain these matched $X$ and $Z$ edges using the assignment method in Ref.~\cite{shutty2024harmony}.   There are other ways of producing representatives, such as relying on an ensemble to stochastically generate matchings in more than one equivalence class~\cite{shutty2024harmony}.  However, the complementary matching also provides the ``complementary gap'', the difference in weight between the two matchings found, which is very predictive of the probability of logical error for a given syndrome~\cite{gidney2023yoked}.  We use the gap to only invoke the ensemble on a small fraction of ``hard'' cases (where the gap is small), which happens sufficiently rarely (and with likelihood that decreases exponentially with code distance when below threshold) that the average runtime of Libra is dictated by complementary matching, and the cost of running an ensemble of 100 decoders becomes insignificant to the average runtime.  Another option would be to use a small ensemble first, where we invoke the larger ensemble if there are least two equivalence classes for the matchings in the first ensemble~\cite{shutty2024harmony}.

There are many ways to choose an ensemble, but we opt for using correlated minimum-weight perfect matching (MWPM)~\cite{fowler2013optimal,Paler2023pipelinedcorrelated,higgott2023improved,shutty2024harmony} with randomly perturbed hyperedge weights.  This is simple to implement (via Stim~\cite{stim}, one can edit DEM files) and is similar to the ensemble in Harmony~\cite{shutty2024harmony}, though not the same.  The perturbation is to multiply the probability $p_i \rightarrow p_i'$ for each error channel by a log-normally distributed random variable: $p_i' = p_i r_i$, where $r_i \sim \mathrm{Lognorm}(0, \sigma^2)$ is equivalent to $r_i = \exp(t_i)$ for $t_i \sim \mathrm{Norm}(0, \sigma^2)$.  For example, if $\sigma = \ln 2$, then the perturbations will have a standard deviation of a factor-of-2, normally distributed in the logarithm of probability.  This will ensure that probabilities are normalized, and zero-probability events (if present) remain zero.  Furthermore, since hyperedge weights are typically the logarithm of probability, this is effectively adding a normally distributed perturbation to hyperedge weights.  To be explicit, we only use these randomly perturbed hypergraph weights to induce an ensemble of correlated-MWPM decoders to give different matching solutions; when we evaluate the weight of a matching or cycle (such as Eqn.~(\ref{eqn:relative_weight}), we use the unperturbed weights.

For the comparison step in Fig.~\ref{fig:libra_architecture}, one could compare the weights of the best solution for each equivalence class (e.g. two for a logical memory experiment) after matching synthesis across the ensemble.  One can also use small-positive cycles found during synthesis to generate multiple configurations in each equivalence class, sum the probabilities for each equivalence class, and compare the probabilities for the equivalence classes, in a manner described below.  Libra is configured to do both (the first is necessary, and the second comes at negligible additional cost), and Section~\ref{sec:simulations} reports the results for decoding the surface code.

We describe here the algorithm currently employed for using the small-positive cycles to estimate probabilities for equivalence classes, though many variations are possible.  As cycles with nonnegative relative weight are discovered, they are stored in a max-heap of fixed size (in the simulations of Section~\ref{sec:simulations}, the size is 30).  When an improving cycle (one with negative relative weight) is found, the heap is cleared, because the stored positive cycles were relative to a hyperedge configuration that has changed with the improving cycle.  Alternatively, one could iterate through the cycles in the heap and only modify or discard as necessary for those overlapping the improving cycle. Because clearing the heap could happen in any synthesis step, Libra currently performs the synthesis procedure twice, where negative cycles are rarely discovered in the second pass, which mostly functions to discover small-positive cycles.

After all matching-synthesis operations are performed, we have a ``best matching'' for each of the equivalence classes being considered by Libra (e.g. two for the memory experiment).  We also have the small-positive cycles that were discovered, from which we can generate many other configurations.  For example, if we store 30 such cycles, we could generate as many as $2^{30} \approx 10^9$ configurations.  The number could be less than $2^{30}$ if the cycles stored are not linearly independent (e.g. if the combination of two cycles is another stored cycle).  Computing all such configurations generated by these cycles would be impractical.  However, at least for surface codes, they tend to be small cycles from disparate parts of the hypergraph, so instead we split the cycles into connected components.  For this step, two cycles are connected if they overlap in at least one edge.  For each connected component of cycles $\{\mathbf{c}_i\}$, we compute a relative probability:
\begin{equation}
    p_r (\{\mathbf{c}_i\}) = \sum_{\mathbf{c} \in \mathcal{G}(\{\mathbf{c}_i\})} p_r(\mathbf{c}),
\end{equation}
where $\mathcal{G}\left(\{\mathbf{c}_i\}\right)$ is the set of all unique cycles generated by $\{\mathbf{c}_i\}$ and the relative probability for a single cycle $\mathbf{c}$ is 
\begin{equation}
    p_r(\mathbf{c}) = p(\mathbf{e} \oplus \mathbf{c}) / p(\mathbf{e}),
\end{equation}
i.e. the ratio of probability of the configuration using the cycle to the ``base'' synthetic matching $\mathbf{e}$.  We emphasize relative, because the ``null'' cycle $\emptyset$ is always an element of any $\mathcal{G}$, corresponding to the synthetic matching itself, and it has a relative probability of 1.  If probability is the logarithm of weight, then relative probability can be computed directly from the relative weight of the cycle, which is much faster than a computation over the entire matching.  Finally, two implementation notes.  First, as we are generating $\mathcal{G}\left(\{\mathbf{c}_i\}\right)$, we may encounter a generating set from the connected component $\{\mathbf{c}_i\}$ that is not linearly independent.  To guard against this, we use a hash table to store the elements of $\mathcal{G}\left(\{\mathbf{c}_i\}\right)$ as they are produced, and skip redundant entries.  An alternative would be to compute an independent vector basis such as by Gaussian elimination over binary vectors.  Second, it is possible that $\mathcal{G}\left(\{\mathbf{c}_i\}\right)$ is very large because it is exponential in the size of $\{\mathbf{c}_i\}$.  For speed, we truncate the sum by only considering a low-order approximation to the exponentially large number of combinations (e.g. use $\{\mathbf{c}_i\}$ as the set to sum over, or all combinations of two elements of $\{\mathbf{c}_i\}$).  In the simulations, we make this approximation if $\vert\{\mathbf{c}_i\}\vert \ge 10$ cycles.  Finally, the estimated probability of the equivalence class is
\begin{equation}
    p(\mathrm{equivalence\;class}) = p(\mathbf{e}) \prod_k p_r^{(k)},
\end{equation}
where $p_r^{(k)}$ is the relative probability for the $k^{\mathrm{th}}$ connected component in the small-positive cycles.

\begin{figure}
  \centering
      \includegraphics[width=0.85\columnwidth]{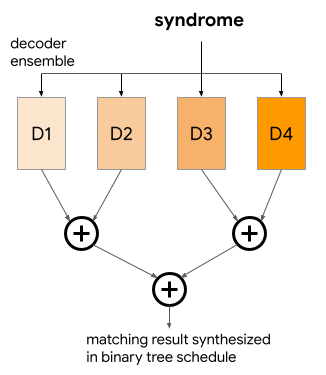}
  \caption{Matching synthesis can be performed by a logarithmic-depth binary tree of synthesis steps.}  
  \label{fig:libra_fusion_tree}
\end{figure}

Another generalization of the procedure is the freedom of how to synthesize members of the ensemble.  Figure~\ref{fig:libra_architecture} illustrates a sequential procedure where the current synthesized matching updates iteratively with each ensemble member.  Alternatively, one could synthesize matchings from the ensemble in a logarithmic-depth binary tree, as shown in Fig.~\ref{fig:libra_fusion_tree}, which could have advantages for parallelization of the algorithm.  Since synthesis can be performed between two matchings, any binary tree with number of leaves equal to ensemble size can describe a synthesis sequence.

\section{The surface code with Pauli errors in the circuit model}
\label{sec:simulations}
For our simulations, we simulate a surface code memory experiment, which consists of initializing in a basis ($X$ or $Z$) by direct initialization of data qubits, repeatedly measuring the stabilizers with a circuit of Clifford gates for some number of cycles, and then measuring the logical qubit in the same basis as initialization, by direct measurement of the data qubits.  Noise in the simulation consists of a depolarizing channel after every operation, with a 1-qubit channel for 1-qubit gates (and reset) or a 2-qubit channel for the 2-qubit gate, which is controlled-Z (CZ) in our circuits.  Measurement error is modeled as a binary symmetric channel on the classical bit (i.e. flip the classical bit with some probability).  The depolarizing channels and measurement error have different probabilities according to the SI-1000 model~\cite{gidney2021honeycomb}, which is representative of error rates in superconducting qubits~\cite{krinner2022,google2023suppressing}.  The weighting also captures features seen in other qubit platforms~\cite{Bluvstein2024,dasilva2024}, such as 1-qubit gates being more reliable than 2-qubit gates, and measurement being less reliable than coherent operations.  In this model, the channel probabilities are only determined by the operation and there is not inhomogeneity in space or time; e.g. all CZ gates have the same depolarizing probability.

\begin{figure*}
  \centering
      \includegraphics[width=0.9\textwidth]{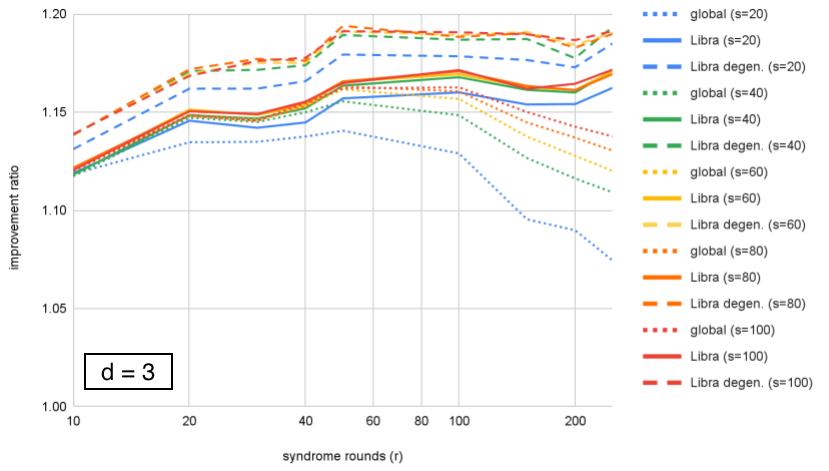}
  \caption{Relative accuracy of the decoders at $d=3$.  Improvement ratio is the ratio of the given decoder's logical error rate to that of correlated MWPM.  For the decoders, global ensembling refers to selecting the single best ensemble member, Libra refers to using matching synthesis, and ``Libra degen'' refers to estimating degeneracy with small-positive cycles.}  
  \label{fig:d3_vs_rounds}
\end{figure*}

\begin{figure*}
  \centering
      \includegraphics[width=0.9\textwidth]{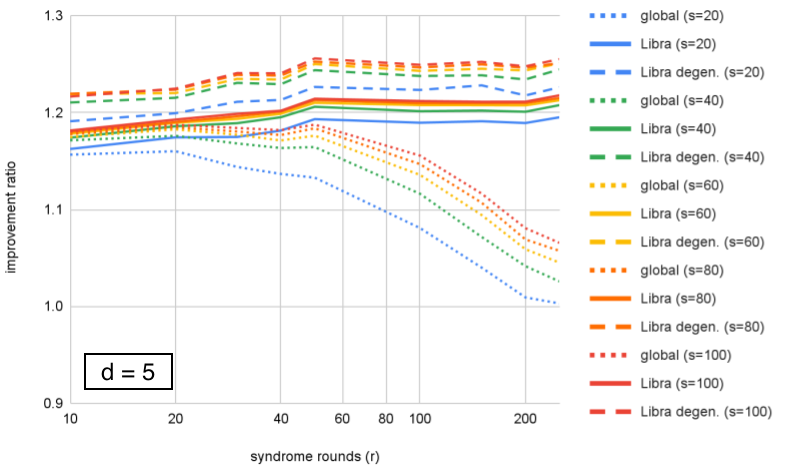}
  \caption{Relative accuracy of the decoders at $d=5$.  As before, improvement ratio is the ratio of the given decoder's logical error rate to that of correlated MWPM.  For the decoders, global ensembling refers to selecting the single best ensemble member, Libra refers to using matching synthesis, and ``Libra degen'' refers to estimating degeneracy with small-positive cycles.}  
  \label{fig:d5_vs_rounds}
\end{figure*}

\begin{figure*}
  \centering
      \includegraphics[width=0.9\textwidth]{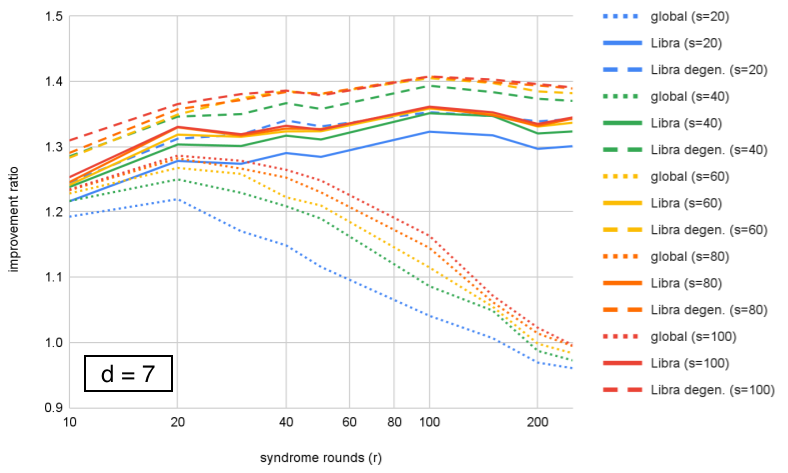}
  \caption{Relative accuracy of the decoders at $d=7$.  As before, improvement ratio is the ratio of the given decoder's logical error rate to that of correlated MWPM.  For the decoders, global ensembling refers to selecting the single best ensemble member, Libra refers to using matching synthesis, and ``Libra degen'' refers to estimating degeneracy with small-positive cycles.}  
  \label{fig:d7_vs_rounds}
\end{figure*}

\begin{figure*}
  \centering
      \includegraphics[width=0.9\textwidth]{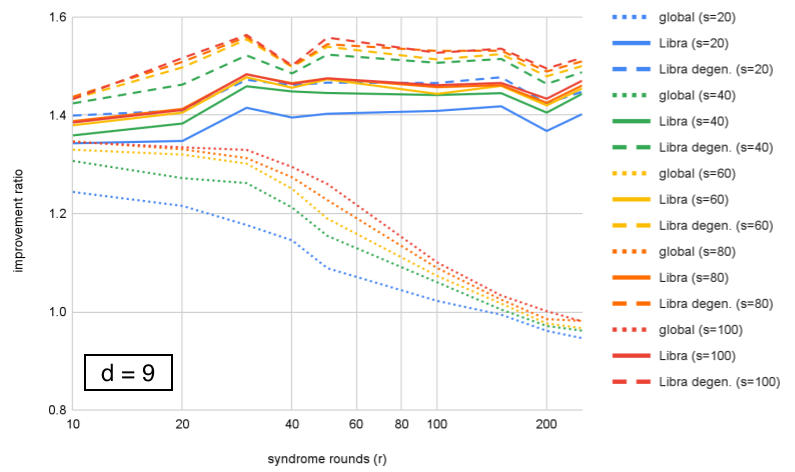}
  \caption{Relative accuracy of the decoders at $d=9$.  As before, improvement ratio is the ratio of the given decoder's logical error rate to that of correlated MWPM.  For the decoders, global ensembling refers to selecting the single best ensemble member, Libra refers to using matching synthesis, and ``Libra degen'' refers to estimating degeneracy with small-positive cycles.}  
  \label{fig:d9_vs_rounds}
\end{figure*}

\begin{figure*}
  \centering
      \includegraphics[width=0.9\textwidth]{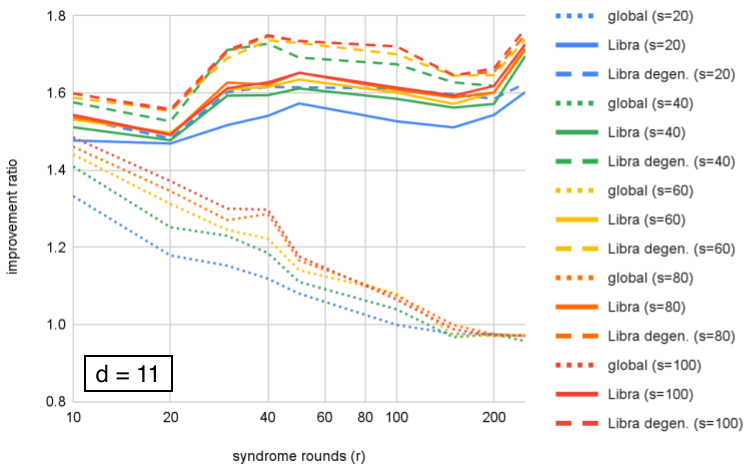}
  \caption{Relative accuracy of the decoders at $d=11$.  As before, improvement ratio is the ratio of the given decoder's logical error rate to that of correlated MWPM.  For the decoders, global ensembling refers to selecting the single best ensemble member, Libra refers to using matching synthesis, and ``Libra degen'' refers to estimating degeneracy with small-positive cycles.}  
  \label{fig:d11_vs_rounds}
\end{figure*}

The simulations here use a simple ensemble for Libra.  As described in Section~\ref{sec:libra_decoder}, each ensemble member is correlated-MWPM~\cite{fowler2013optimal,Paler2023pipelinedcorrelated,higgott2023improved,shutty2024harmony} configured with the hypergraph for the circuit (i.e. DEM file~\cite{stim}), but with the error probabilities randomly perturbed by $\sim \mathrm{Lognorm}(0, \sigma^2)$.  Half of the members have $\sigma = \ln 2$ and the other half $\sigma = \ln 4$.  In testing, we found such an inhomogeneous ensemble to lead to a small improvement over a homogeneous ensemble, which may be due to different values of $\sigma$ enabling the ensemble to discover a greater variety of cycles.  Future work will explore alternative ways to construct an ensemble.  Libra also includes a complementary matcher (Fig.~\ref{fig:libra_architecture}) with unperturbed hypergraph, as described in Section~\ref{sec:libra_decoder}.

The simulations are a surface-code memory experiment, in the logical $Z$ basis, using CZ and Hadamard gates, with SI-1000 depolarizing model with parameter $p = 2\mathrm{e-}3$.  The parameters swept in the simulations are code distance ($d$), syndrome rounds ($r$), and ensemble size ($s$):
\begin{itemize}
    \item $d \in \{3, 5, 7, 9, 11\}$,
    \item $r \in \{10, 20, 30, 40, 50, 100, 150, 200, 250\}$,
    \item $s \in \{20, 40, 60, 80, 100\}$.
\end{itemize}
For every combination of $(d, r, s)$ we perform correlated MWPM as a baseline, and complementary matching to determine a complementary gap and initialize two equivalence classes in Libra.  If the complementary gap is less than 20 dB (i.e. ratio of probability between the two matchings is $<100$), then we run the ensemble; this conditional execution~\cite{delfosse2020hierarchical,shutty2024harmony} significantly reduces computation time since the fraction of problems where the ensemble is run scales with the logical error rate~\cite{gidney2023yoked}, decreasing with distance.  With the ensemble, we produce three decoder predictions: 
\begin{itemize}
    \item ``global'' ensembling, which is using the minimum-total-weight error configuration found by an ensemble member~\cite{shutty2024harmony}; 
    \item Libra using matching synthesis to make improved configurations;
    \item Libra ``degeneracy'' using small-positive cycles to estimate the probability of equivalence classes.
\end{itemize}
Each of these three ensembled methods use exactly the same ensemble (described above), so they demonstrate the effects of matching synthesis and using small-positive cycles.  However, it is important to note that different ensembles (e.g. distributions of weight perturbations for matching) may be better suited to each of the ensembling methods listed above.  The size of the heap for smallest-relative-weight cycles is 30 in all cases.  When the ensemble is not run because the complementary gap is large, all decoders default to the prediction of correlated MWPM.

We use correlated MWPM as the baseline decoder, which yields error suppression factor~\cite{google2023suppressing} $\Lambda \approx 3.6$ for SI-1000($p = 2\mathrm{e-}3$).  We quantify accuracy as the ratio of improvement relative to the correlated-MWPM baseline:
\begin{equation}
    \mathrm{improvement \; ratio}(\mathrm{decoder}) \; = \; \frac{\epsilon_{\mathrm{MWPM}}}{\epsilon_{\mathrm{decoder}}}.
\end{equation}
We calculate logical error rate for a memory experiment of $r$ rounds as
\begin{equation}
    \epsilon = \frac{1}{2}\left(1 - \left(1 - 2\frac{n_f}{N}\right)^{(1/r)}\right),
\end{equation}
where $N$ is total number of samples and $n_f$ is number of simulated failures.  For all values of $(d, r, s)$, we simulate to at least 1000 failures for every decoder.  For a fixed $(d, r)$ combination, all decoders for all values of $s$ see the same sampled shots.

\begin{figure*}
  \centering
      \includegraphics[width=0.9\textwidth]{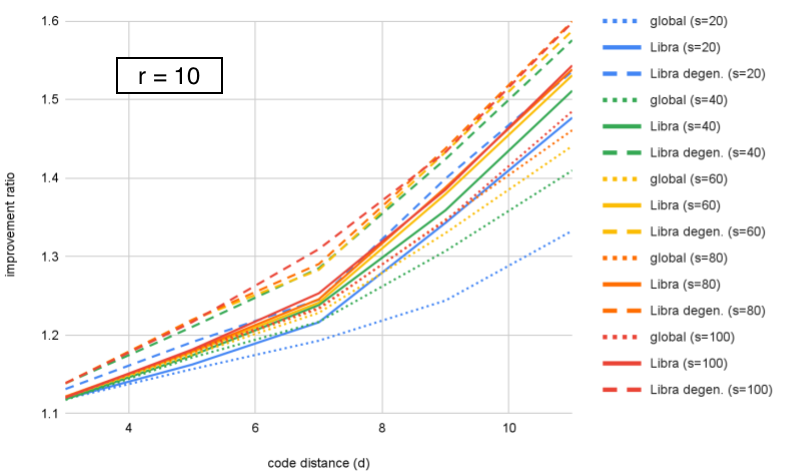}
  \caption{Relative accuracy of the decoders as a function of distance for $r=10$.  As before, improvement ratio is the ratio of the given decoder's logical error rate to that of correlated MWPM.  For the decoders, global ensembling refers to selecting the single best ensemble member, Libra refers to using matching synthesis, and ``Libra degen'' refers to estimating degeneracy with small-positive cycles.}  
  \label{fig:r10_vs_d}
\end{figure*}

\begin{figure*}
  \centering
      \includegraphics[width=0.9\textwidth]{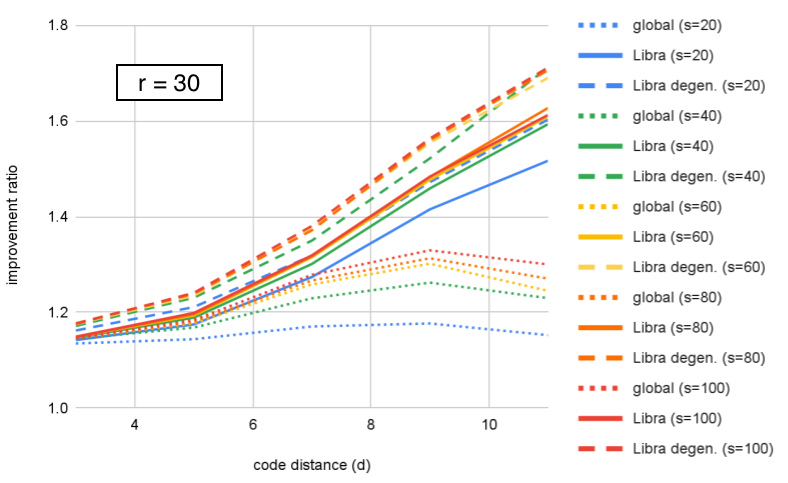}
  \caption{Relative accuracy of the decoders as a function of distance for $r=30$.  As before, improvement ratio is the ratio of the given decoder's logical error rate to that of correlated MWPM.  For the decoders, global ensembling refers to selecting the single best ensemble member, Libra refers to using matching synthesis, and ``Libra degen'' refers to estimating degeneracy with small-positive cycles.}  
  \label{fig:r30_vs_d}
\end{figure*}

\begin{figure*}
  \centering
      \includegraphics[width=0.9\textwidth]{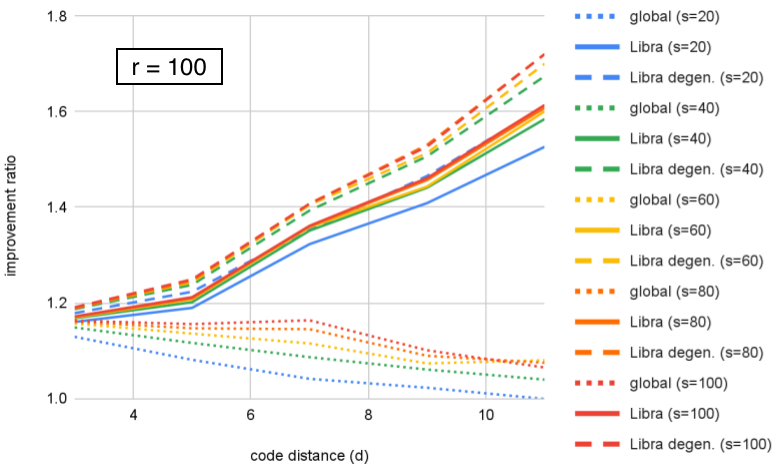}
  \caption{Relative accuracy of the decoders as a function of distance for $r=100$.  As before, improvement ratio is the ratio of the given decoder's logical error rate to that of correlated MWPM.  For the decoders, global ensembling refers to selecting the single best ensemble member, Libra refers to using matching synthesis, and ``Libra degen'' refers to estimating degeneracy with small-positive cycles.}  
  \label{fig:r100_vs_d}
\end{figure*}

\begin{table*}[]
    \centering
    \begin{tabular}{l c c c}
       \textbf{Decoder}  &  $\mathbf{\Lambda_{3,11}}$ & $\mathbf{\Lambda_{7,11}}$ & $\mathbf{\Lambda_{9,11}}$\\ \hline
       MWPM-corr & 3.68 & 3.64 & 3.52 \\ 
       global ens. \\
       \; ens=20 & 3.70 & 3.61 & 3.45 \\ 
       \; ens=40 & 3.75 & 3.64 & 3.43 \\
       \; ens=60 & 3.76 & 3.62 & 3.37 \\
       \; ens=80 & 3.78 & 3.65 & 3.41 \\
       \; ens=100 & 3.80 & 3.67 & 3.44 \\ 
       Libra \\ 
       \; ens=20 & 3.95 & 3.97 & 3.78 \\
       \; ens=40 & 4.00 & 4.03 & 3.85 \\ 
       \; ens=60 & 4.01 & 4.03 & 3.84 \\
       \; ens=80 & 4.02 & 4.04 & 3.86 \\
       \; ens=100 & 4.01 & 4.02 & 3.83 \\
       Libra degen. \\
       \; ens=20 & 3.99 & 4.01 & 3.84 \\
       \; ens=40 & 4.05 & 4.10 & 3.96 \\
       \; ens=60 & 4.03 & 4.04 & 3.83 \\
       \; ens=80 & 4.04 & 4.06 & 3.85 \\
       \; ens=100 & 4.05 & 4.05 & 3.85 \\
    \end{tabular}
    \caption{Values for error-suppression ratio $\Lambda$ calculated at $r=30$ for various decoders.  The ensemble size is reported as ``ens=20'', \emph{etc}.  This table computes $\Lambda$ in three ways, which yield slightly differing results: for logical error rate $\epsilon_d$ at code distance $d$, $\Lambda_{3,11} = \left(\epsilon_3 / \epsilon_{11}\right)^{1/4}$, $\Lambda_{7,11} = \left(\epsilon_7 / \epsilon_{11}\right)^{1/2}$, and $\Lambda_{9,11} = \left(\epsilon_9 / \epsilon_{11}\right)$.}
    \label{tab:lambda}
\end{table*}

We compare performance of the ensembled decoders at each $d$, showing improvement over correlated-MWPM as a function of $r$.  In Figs.~\ref{fig:d3_vs_rounds}--\ref{fig:d9_vs_rounds}, we see two effects.  First, Libra maintains performance (or improves slightly) with number of rounds, while global ensembling decreases; the onset of the decrease becomes earlier with increasing $d$, showing that both $d$ and $r$ affect performance of global ensembling.  Second, including degeneracy by using small-positive cycles to compute probabilities for equivalence classes shows an additional modest benefit over using the best single matching.

We can look at the same data organized differently, by grouping by fixed $r$ value and plotting improvement as a function of distance.  This is plotted for $r \in \{10, 30, 100\}$ in Figures~\ref{fig:r10_vs_d}--\ref{fig:r100_vs_d}.  We can see that Libra shows increasing improvement ratio with distance for all cases.  For global ensembling, there is improvement for $r=10$ with $d$, some slowing of improvement for $r=30$, and a decrease with distance for $r=100$.  Section~\ref{sec:interpretation} proposes an explanation for this.  If we take $r=30$ as an example, we see that both Libra and Libra-degen. appear to saturate in performance around ensemble size $s=60$.

Figures~\ref{fig:r10_vs_d}--\ref{fig:r100_vs_d} show that the improvement ratio is increasing with $d$ in both versions of Libra.  The rate of improvement can be quantified with $\Lambda$, and values for the different decoders are calculated in Table~\ref{tab:lambda} for $r = 30$.  If we take ensemble size $s = 100$ for example, we find that $\Lambda_{7,11}$ increases from 3.64 to 4.02 (+10\%) for Libra  or to 4.05 (+11\%) when including ``degeneracy'' from small-positive cycles.  

However, we remark that $\Lambda$ is not the only figure of merit -- being a ratio of logical error rates, it increases if the numerator (a smaller code) has higher logical error rate.  For example, in Table~\ref{tab:lambda}, we see that at $s = 80$, Libra has $\Lambda_{9,11}$ = 3.86 and Libra-degen. has $\Lambda_{9,11}$ = 3.84 (as a reminder, these are decoding exactly the same samples of a memory experiment).  Conversely, from Fig.~\ref{fig:r30_vs_d}, we see that Libra-degen. achieves higher improvement ratio (hence lower logical error rate) for $d=9$ and $d=11$.  

We see overall that Libra is improving by $\Lambda$ by about 10\% for the range of parameters studied here, but generalizing to higher distance or different error rates is a matter for future study.

\section{Interpretation of Libra}
\label{sec:interpretation}
In this section, we conjecture on the mechanism that enables Libra to work well for surface-code decoding.  The key concept in Libra is identifying improving cycles.  How well this works will depend on the structure of the weighted hypergraph and the nature of the ensemble.  In some informal sense, the ensemble needs some ``entropy'' (yielding a diversity of solutions) but not too much (for a local neighborhood of the hypergraph, one of the ensemble members needs to find a good local solution).  

In our simulations with the surface code (Section~\ref{sec:simulations}), we found that Libra improved accuracy over correlated MWPM~\cite{fowler2013optimal,Paler2023pipelinedcorrelated,higgott2023improved,shutty2024harmony}.  We attribute this to 3D locality of the hypergraph and the fact that the decoders in the ensemble produce a diversity of reasonably good solutions.  The surface code is a topological code~\cite{kitaev1998surfacecode,Terhal2015} and the error model is local to each operation, so the hypergraph is local in three dimensions (every hyperedge is contained within a ball of some finite radius, independent of the code distance).  Moreover, correlated MWPM is a ``pretty good'' decoder for the surface code, in the sense that high-accuracy decoders have demonstrated only modestly better performance ~\cite{google2023suppressing,bausch2023learning,shutty2024harmony}, meaning $\Lambda$ increases by about $10\%$.  Hence the members of the ensemble tend to produce solutions that are good in a local neighborhood.  We speculate that Libra will show benefit with other topological codes, such as color codes~\cite{landahl2011fault,gidney2023chromobius}, provided suitably ``good'' approximate decoders for the ensemble.  Similarly, we expect Libra to generalize naturally to surface-code logical operations~\cite{horsman2012, fowler2012largescale, fowler2018,gidney2024ybasis}, since these also have the properties of the error hypergraph being 3D-local and being amenable to matching.

We give some intuition for how Libra improves accuracy of decoding, though we preface that this story will be studied more carefully in future work.  When operating below the threshold for error correction, which is the domain of interest for decoding, errors tend to be sparsely distributed~\cite{dennis2002topological,higgott2023sparse}.  In matching synthesis, this leads to small, local cycles.  

\begin{figure}
  \centering
      \includegraphics[width=0.85\columnwidth]{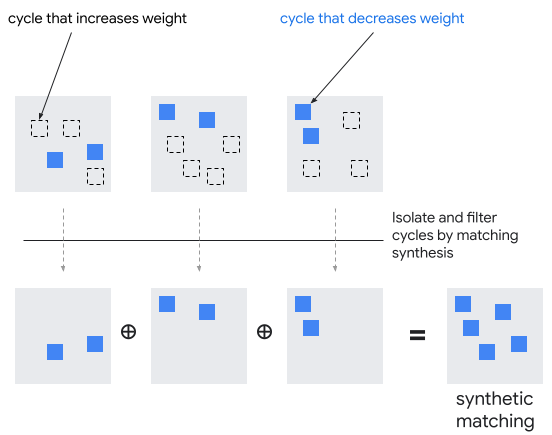}
  \caption{Matching synthesis, and the Libra decoder, filter cycles by relative weight.  Each gray box is a matching solution from the ensemble, and the smaller dashed or solid-blue boxes represent cycles of positive or negative relative weight, referenced to a baseline solution (in the text, we use correlated MWPM as a baseline). The synthesized matching only incorporates the improving cycles.}  
  \label{fig:libra_cycle_filter}
\end{figure}

As a frame of reference, let us use the matching result produced by correlated MWPM as the baseline configuration of errors, and we can represent any other configuration as the collection of cycles and logical operators produced by symmetric difference between the two.  We observe that with randomly perturbed error probabilities, the cycles discovered in matching synthesis have an average weight that is positive.  This means that the majority are not improving cycles.  Hence, if we were to take whole matchings from the ensemble and only consider their total weight~\cite{shutty2024harmony}, then we find a better solution than the baseline only when an ensemble member gets lucky enough to sample enough improving cycles to be in the tail of the distribution.  In contrast, Libra is able to select the improving cycles individually, shown in Fig.~\ref{fig:libra_cycle_filter}.

\begin{figure}
  \centering
      \includegraphics[width=0.85\columnwidth]{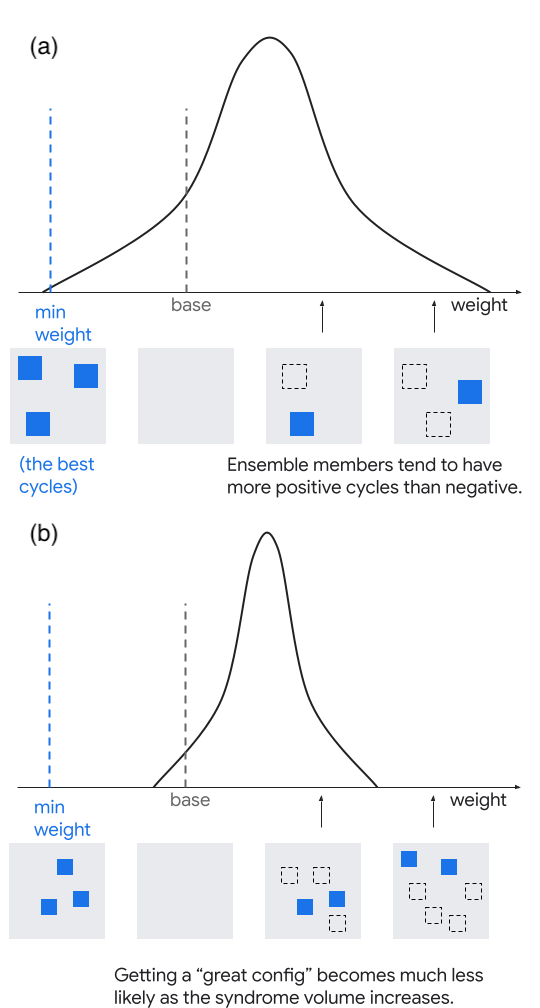}
  \caption{Conceptual explanation for how ensembling with total weight of an error configuration converges to the average performance of an ensemble member as size of the error hypergraph increase from (a) to (b), with a distribution of cycles that have average positive relative weight.  The conceptual representation of positive- or negative-weight cycles with dashed or solid-blue boxes, respectively, is the same as Fig.~\ref{fig:libra_cycle_filter}.  The label ``base'' refers to a baseline decoder which is the reference for the relative weight of cycles -- for example, correlated MWPM.  The label ``min weight'' is the optimal solution, shown here only for illustration along the ``weight'' axis.}  
  \label{fig:cycle_distr_concept}
\end{figure}

One might expect that there will be some density distribution of cycles that is independent of code distance (i.e. diameter of the hypergraph), owing to the local structure of the hypergraph described above.  We caution that this is a conjecture, since we have not yet studied this carefully.  This distribution could also be correlated with other observed values, such as there being more improving cycles when the complementary gap is small.  

If the relative weight of each cycle can be modeled as a real-valued random variable with positive average value, then the ``global'' difference in weight between an ensemble member and the correlated-MWPM baseline, which is the sum over the cycle relative weights, will also have positive average value.  Moreover, by the Central Limit Theorem, the ratio of the average and standard deviation for this global weight difference will shrink if the number of cycles increases (i.e. sample mean converges to distribution mean).  We conjecture that the number of cycles will increase with size of the hypergraph (i.e. volume of syndrome, which in a memory experiment is $\sim d^2 r$ for code distance $d$ and number of syndrome cycles $r$), as depicted in Fig.~\ref{fig:cycle_distr_concept}.  In such circumstances, using the global lowest-weight configuration in the ensemble as the solution~\cite{shutty2024harmony} will require the size of the ensemble to increase with the syndrome volume, in order to sample from the tail of the distribution for global weight.  In contrast, Libra would be insensitive to this globally imposed Central Limit Theorem because it can select the best local cycles individually.  The data in Section~\ref{sec:simulations} supports this interpretation, where we see that Libra performance is insensitive to $d$ and $r$, whereas global-weight ensembling is sensitive to both and recovers by increasing the size of the ensemble.  However, we do not propose a functional form for this dependence.  
Techniques like windowing~\cite{skoric2023,tan2023scalable} or graph partitioning~\cite{wu2023fusionblossom} are important for parallelization of decoding, as well as streaming for real-time operation.  If ensembling is applied to increase accuracy, windowing and/or partitioning the hypergraph would limit the size of the syndrome volume in the cut-out region, mitigating the volume-dependent effects described above.  However, we expect that the minimum size of the syndrome volume will scale $\sim d^3$ for the surface code, since logical operators of length $\sim d$ can be oriented in any of the three cardinal directions in 3D.  Hence, these considerations become important as quantum error correction scales to large $d$.

\section{Generalizations to other problems}
This work focuses on solving a particular variant of MWHPM (Section~\ref{sec:MWHPM_def}) that is relevant to decoding quantum codes, but the methods appear to be more general.  If one is performing constrained optimization where solutions can be decomposed into pieces and evaluated individually, such as Eqn.~(\ref{eqn:weight_optimization}), then it could be possible to synthesize two distinct solutions and keep the best pieces.  In contrast to local search in stochastic algorithms~\cite{lin1965,lin1973,hutter2014}, this could avoid becoming trapped in local minima~\cite{martin1991}.

Using small-positive cycles (from Section~\ref{sec:matching_synthesis}) as generators of multiple ``good'' solutions could also be of interest in optimization.  For example, suppose one is optimizing routing in a distribution network, and that there is a formulation of this problem to which matching synthesis or a generalization thereof can be applied.  An optimal solution that has no small cycles could be interpreted as ``brittle'', because if the real-world circumstances that inform the cost function change, the single solution could become a poor one.  In contrast, if there is a solution with many small-positive cycles, it means there are many almost-best solutions that could be generated.  In the example, we might say this is a ``robust'' solution because there are many options to make changes to routing, while still being a pretty good solution.  Whether these conjectures can be applied usefully in optimization problems is left for future work.

\section{Discussion}
We have presented a method for high-accuracy decoding of surface codes called matching synthesis, and demonstrated improvements in accuracy with a decoder called Libra.  We showed that the method makes efficient use of ensembling, reaching saturation around an ensemble size of 60 when using correlated MWPM.  Our interpretation of the algorithm and these results is that each matching in the ensemble can be characterized as differing from the optimal solution by some random distribution of cycles on the hypergraph.  If the hypergraph has ``local structure'' whereby these cycles tend to be confined to local neighborhoods, then matching synthesis quickly approaches optimality when the ensemble finds a good solution in each neighborhood with high probability, which is more favorable than requiring an ensemble member to find a good global solution.

A question not answered here is how close Libra comes to optimal decoding, which we will investigate in future work. In the case of surface codes, while it is theoretically interesting to wonder how close one can get to optimal accuracy with a computationally efficient algorithm, what is the practical benefit?  If Libra (or any other decoder) increases $\Lambda$ by 10\%, then the practical benefit is to reduce the code distance $d$ required to achieve a target logical error rate $\epsilon$ according to the approximate formula $\epsilon = 0.1 \Lambda ^{-(d+1)/2}$~\cite{fowler2012largescale,jones2012arch}.  For example, if the target is $\epsilon = 1\mathrm{e-}12$ per syndrome round and the baseline decoder (e.g. correlated MWPM) achieves $\Lambda = 4$, then the benefit of an improved decoder with $\Lambda = 4.4$ is to reduce required code distance from $d=37$ to $d=35$, reducing qubit overhead by about $1 - (35/37)^2 \approx 11\%$.  

Besides chasing diminishing returns in accuracy, another avenue to applying these results is real-time decoding.  An advantage inherent in ensembled decoding~\cite{shutty2024harmony} is the ability to do most of the decoding in parallel.  Furthermore, matching synthesis with a logarithmic-depth tree lends itself to short-depth computation. Instead of making an ensemble out of the already-pretty-good correlated MWPM decoder, one could instead use simpler algorithms for the ensemble like uncorrelated matching~\cite{dennis2002topological}, Union-Find~\cite{delfosse2021unionfind}, clustering decoders~\cite{bravyi2013cluster}, or renormalization group decoders ~\cite{duclos2010a,duclos2010b}.  Being less accurate, these might require a larger ensemble; however, if they can run in parallel on massively parallel hardware (e.g. FPGAs or GPUs) and be synthesized in a logarithmic-depth tree, the execution depth might actually be shorter than MWPM.

Finally, the matching synthesis method is not restricted to surface codes.  We expect that the technique will translate well to other topological codes, such as color codes~\cite{bombin2008,landahl2011fault,delfosse2014,sahay2022mobius,benhemou2023colorcode,kubica2023restriction,gidney2023chromobius}.  Whether it offers a benefit for other families of codes, such as quantum LDPC codes ~\cite{kovalev2013ldpc,breuckmann2021ldpc,bravyi2024ldpc}, is not clear at this time.  We leave these investigations to future work.

\begin{acknowledgments}
We thank Michael Newman, Dave Bacon, Oscar Higgott, and Noah Shutty for feedback on the matching synthesis procedure.
\end{acknowledgments}

\bibliography{references}
\end{document}